\documentclass[a4paper, oneside, twocolumn, notitlepage, 10pt]{extarticle_ecoc}
\usepackage{ecoc}

\usepackage{cleveref}
\crefname{equation}{Eq.}{Eq.} 
\crefrangelabelformat{equation}{(#3#1#4)--(#5#2#6)}
\crefname{figure}{Fig.}{Fig.}
\crefname{table}{Table}{Table}

\begin{document}
\selectlanguage{english}    


\title{End-to-end optimization of directly modulated laser links using chirp-aware modeling}%


\author{
    Sergio Hernandez F.\textsuperscript{(1)}, Christophe Peucheret \textsuperscript{(2)},
    Francesco Da Ros\textsuperscript{(1)}, Darko Zibar\textsuperscript{(1)}
}

\maketitle                  


\begin{strip}
    \begin{author_descr}

        \textsuperscript{(1)} Department of Electrical and Photonics Engineering, Technical University of Denmark, 
        \textcolor{blue}{\uline{shefe@dtu.dk}}

        \textsuperscript{(2)} Univ Rennes, CNRS, FOTON - UMR6082, 22305 Lannion, France.

    \end{author_descr}
\end{strip}

\renewcommand\footnotemark{}
\renewcommand\footnoterule{}


\begin{strip}
    \begin{ecoc_abstract}
        The rate and reach of directly-modulated laser links is often limited by the interplay between chirp and fiber chromatic dispersion. We address this by optimizing the transmitter, receiver, bias and peak-to-peak current to the laser jointly. Our approach outperforms Volterra post-equalization at various symbol rates.   
    \end{ecoc_abstract}
\end{strip}


\section{Introduction}
The low cost, low power consumption and reduced form factor of directly modulated lasers (DMLs) make them a competitive choice in short-reach intensity modulation / direct detection (IM/DD) optical links \cite{9720189}. Nonetheless, the DML dynamics induce impairments such as bandwidth limitation, nonlinear waveform distortion and frequency chirping. The bandwidth limitation introduces intersymbol interference (ISI) as the symbol rate $R_s$ is increased, degrading the error performance of the link. The chirp interacts with the fiber chromatic dispersion (CD), potentially causing additional ISI. This makes frequency chirp one of the main limiting factors in the transmission of DML-generated signals, especially at high symbol rates \cite{9874980}. 

Traditional impairment compensation in DML-based systems has relied on receiver-only equalization (EQ) \cite{REZA2020102322} or alternating optimization of digital pre-distortion (DPD) and EQ filters \cite{Kottke:17}, where only one of the filters is optimized at a time. Such approaches simplify the filter optimization, but can lead to suboptimal solutions compared to joint optimization techniques \cite{8553260}. Recent approaches have relied on end-to-end (E2E) learning to replace certain functions at the transmitter (TX) and receiver (RX) by a neural network. This has allowed to learn geometrical constellation shaping (GCS) and symbol decoding simultaneously \cite{Srinivasan:22}. The optimization is done through the use of a differentiable DML surrogate model, that allows the propagation of gradients between TX and RX and enables the use of numerical optimizers. However, such models have so far only been trained based on optical power sequences and they consequently disregard the frequency chirp of the laser \cite{10382548}. This restricts their use to back-to-back configurations or scenarios with low dispersion.

In this paper, we propose a chirp-aware DML link optimization using E2E impairment compensation techniques. The data-driven surrogate modeling is based on an extension of our previously proposed methodology employing convolutional attention transformers (CATs) \cite{10382548}, where we now also estimate the instantaneous phase of the optical field generated by the laser. This enables to take the interplay between chirp and fiber dispersion into account in the link optimization. 
The E2E impairment compensation is based on an autoencoder (AE). The AE is trained to jointly optimize GCS and DPD on the TX side, and EQ and symbol detection on the RX side. Additionally, the bias current $I_\mathrm{bias}$ and peak-to-peak current $I_\mathrm{pp}$ driving the laser are also controlled by the AE, since those parameters critically impact the amplitude and phase dynamics of the DML. This allows to optimize the TX-side waveform generation for a certain accumulated dispersion value. The performance of the AE is benchmarked against two RX-only equalization schemes: one based on a linear finite impulse response equalizer (FIR EQ), and a second one based on a second-order Volterra nonlinear equalizer (VNLE). The AE is able to outperform the RX-side equalizers at every tested symbol rate $R_s$ for the same total number of taps employed. 
\section{Simulation setup}
\begin{figure*}[t]
    \centering
    \includegraphics[width=\textwidth]{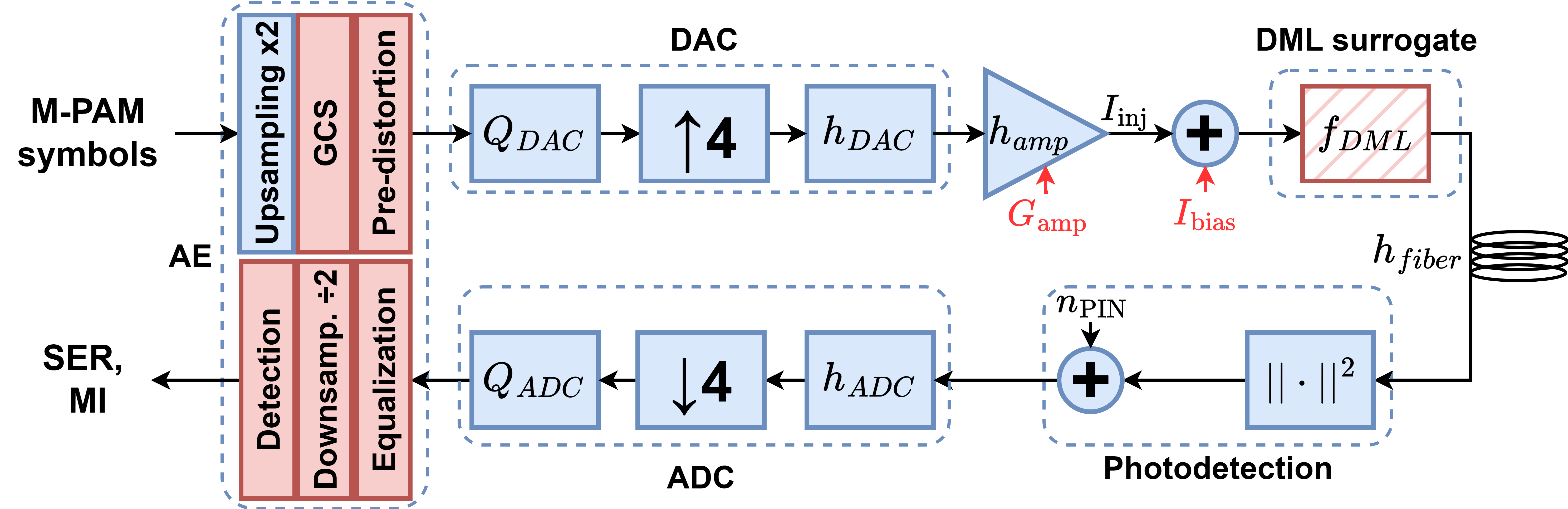}
    \caption{Block diagram of the proposed simulated setup. The optimized elements are highlighted in red. The DML surrogate model $f_\mathrm{DML}$ is trained separately and kept fixed during link optimization.}
    \label{fig:setup}
\end{figure*}

The link simulation aims to replicate an IM/DD link using a DML as laser source at a central wavelength $\lambda_0 = 1286$ nm and a transmission distance of 2 km over standard single mode fiber (SSMF). The block diagram of the system under investigation is shown in \cref{fig:setup}. The encoder (TX-side) of the AE maps 4-level pulse amplitude modulation (4PAM) symbols to a temporal sequence, sent to the digital to analog converter (DAC) for transmission. The DAC model includes a 5-bit resolution quantization function $Q_{DAC}$, approximated with a second-order Fourier series, 4-SpS upsampling to simulate the analog domain, and a low-pass filter $h_{DAC}$ modeled as a second-order super-Gaussian FIR with cut-off frequency of $0.85 R_s$. The signal is then amplified through the tunable gain $G_\mathrm{amp}$ to obtain the desired $I_\mathrm{pp}$. The bandwidth limitation of the amplifier is not considered, so $h_\mathrm{amp}$ is a 1-tap FIR filter. The DML response $f_\mathrm{DML}$ maps the injection current $I_\mathrm{inj}$ and $I_\mathrm{bias}$ driving the laser into the instantaneous optical field. $f_\mathrm{DML}$ is approximated by training the CAT model and keeping its parameters fixed during link optimization. The $h_{fiber}$ is the impulse response of the dispersive fiber with an accumulated normal dispersion $D = -4 \; \mathrm{ps}/ \mathrm{nm}$ \cite{8207825}. The photodetection (PD) model consist in square-law detection and additive Gaussian noise $n_\mathrm{PIN}$. The noise power (variance of $n_\mathrm{PIN}$) is adjusted to obtain a maximum electrical signal-to-noise ratio (SNR) of 20 dB, corresponding to $I_\mathrm{pp} = 20$~mA. The SNR calculation is performed based on the average received radio-frequency power $P_{RF}$, that corresponds to the standard deviation of the received optical power waveform. The analog-to-digital converter (ADC) model is similar to the DAC model in reverse order, substituting the upsampling by downsampling. Finally, the AE decoder (RX-side) maps the received temporal sequence to symbol probabilities, allowing the calculation of symbol error rate (SER) and mutual information (MI).

\begin{figure}[b]
    \centering
    \includegraphics[width=\linewidth]{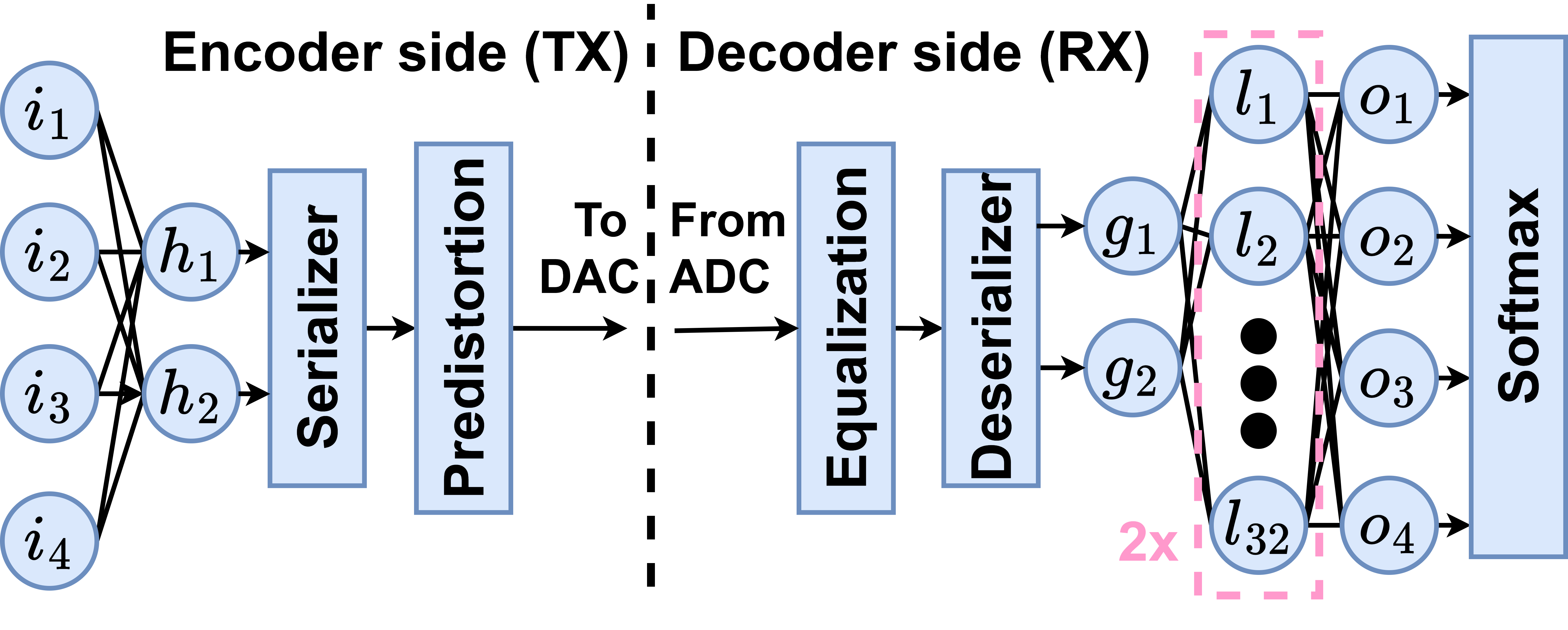}
    \caption{Structure of the selected AE on its encoder (left) and decoder (right) side}
    \label{fig:ae}
\end{figure}
The proposed chirp-aware CAT-based DML modeling is performed in a data-driven fashion, using the laser rate equations as source of data. 
The input sequences to the model are generated using 4PAM symbols and randomized pulse shaping \cite{10382548}. 
Once the input sequences are obtained, they are used as input to a Runge-Kutta (RK4,5) ordinary differential equation (ODE) solver of the rate equations. This allows to obtain the output photon density $S(t)$ and carrier density $N(t)$ sequences corresponding to the applied current sequences, and thus the amplitude and phase of the optical field through finite difference approximations. The output sequences are used as targets to the surrogate model, forcing it to predict both $S(t)$ and $N(t)$ simultaneously. The utilized loss function is the normalized root mean squared error (NRMSE) between predicted and target sequences. The CAT surrogate NRMSE must be divided into $S(t)$ loss and $N(t)$ loss, given the range differences between the two quantities.  

\begin{figure}[b!]
    \centering
    \includegraphics[width=\linewidth]{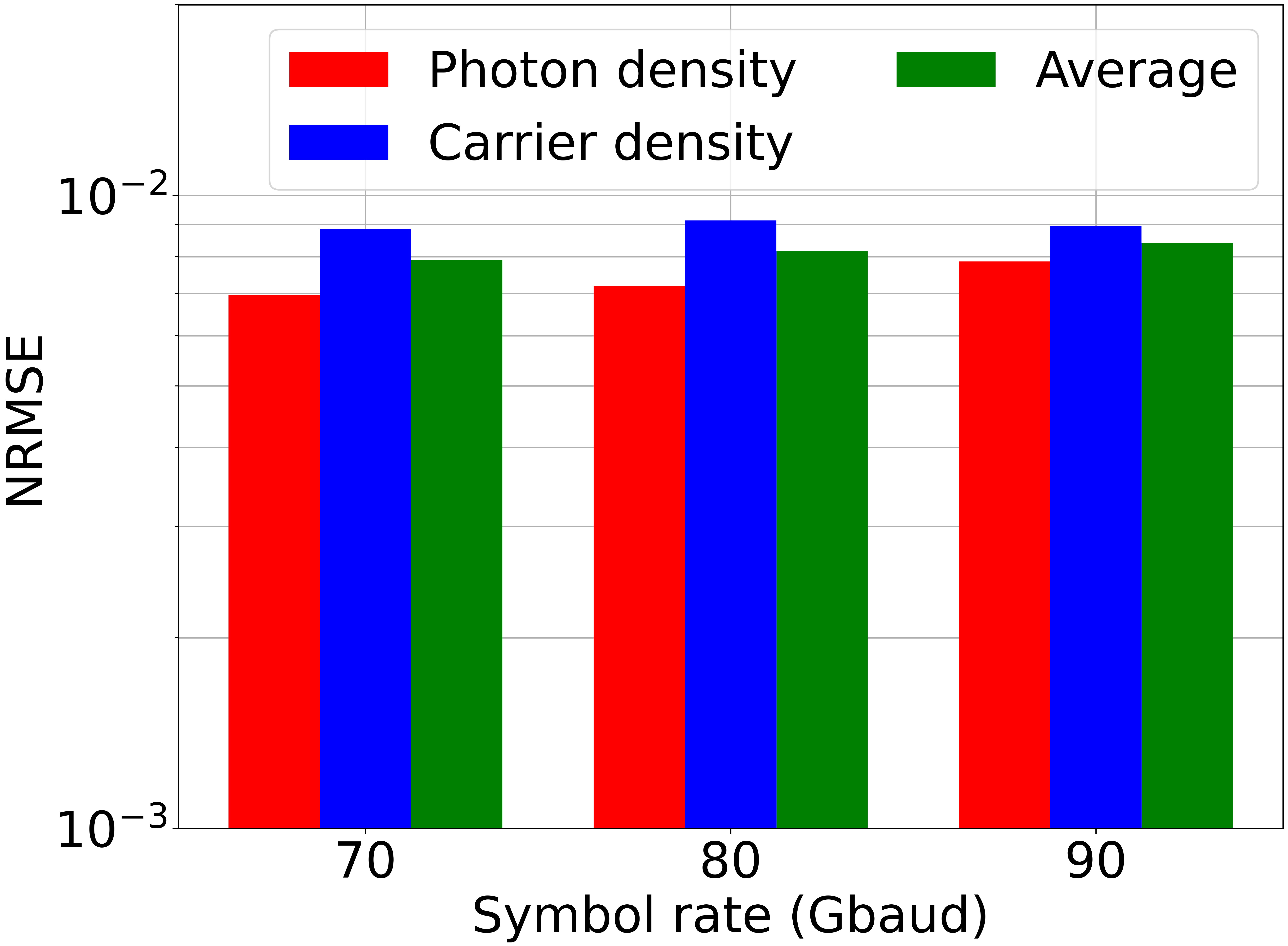}
    \caption{Test NRMSE performance of the chirp-aware surrogate DML model}
    \label{fig:surr}
\end{figure}

The AE structure is shown in \cref{fig:ae}. The encoder maps each one-hot encoded 4PAM symbol $i_n$ linearly into 2-SpS pulse shape $h_n$, implementing GCS. The pulse shapes are then serialized in order to obtain a temporal sequence. An FIR DPD filter is applied to the temporal sequence to pre-compensate for ISI. $I_\mathrm{pp} \in [0.1,20]$ mA and $I_\mathrm{bias} \in [15,30]$ mA are implemented as optimizable parameters, allowing the AE model to infer their optimal value based on the channel conditions. The decoder takes the received quantized signal and performs linear FIR equalization. A deserializer converts the equalized temporal sequence back into 2-SpS waveforms $g_n$. Two hidden layers of 32 neurons $l_n$ with rectified linear (ReLU) activation and a 4-neuron output layer $o_n$ with softmax activation converts each $g_n$ into symbol probabilities. The categorical cross-entropy (CE) between TX and RX symbols was used as loss function for the AE training.

\begin{figure*}[t!]
    \centering
    \includegraphics[width=\textwidth]{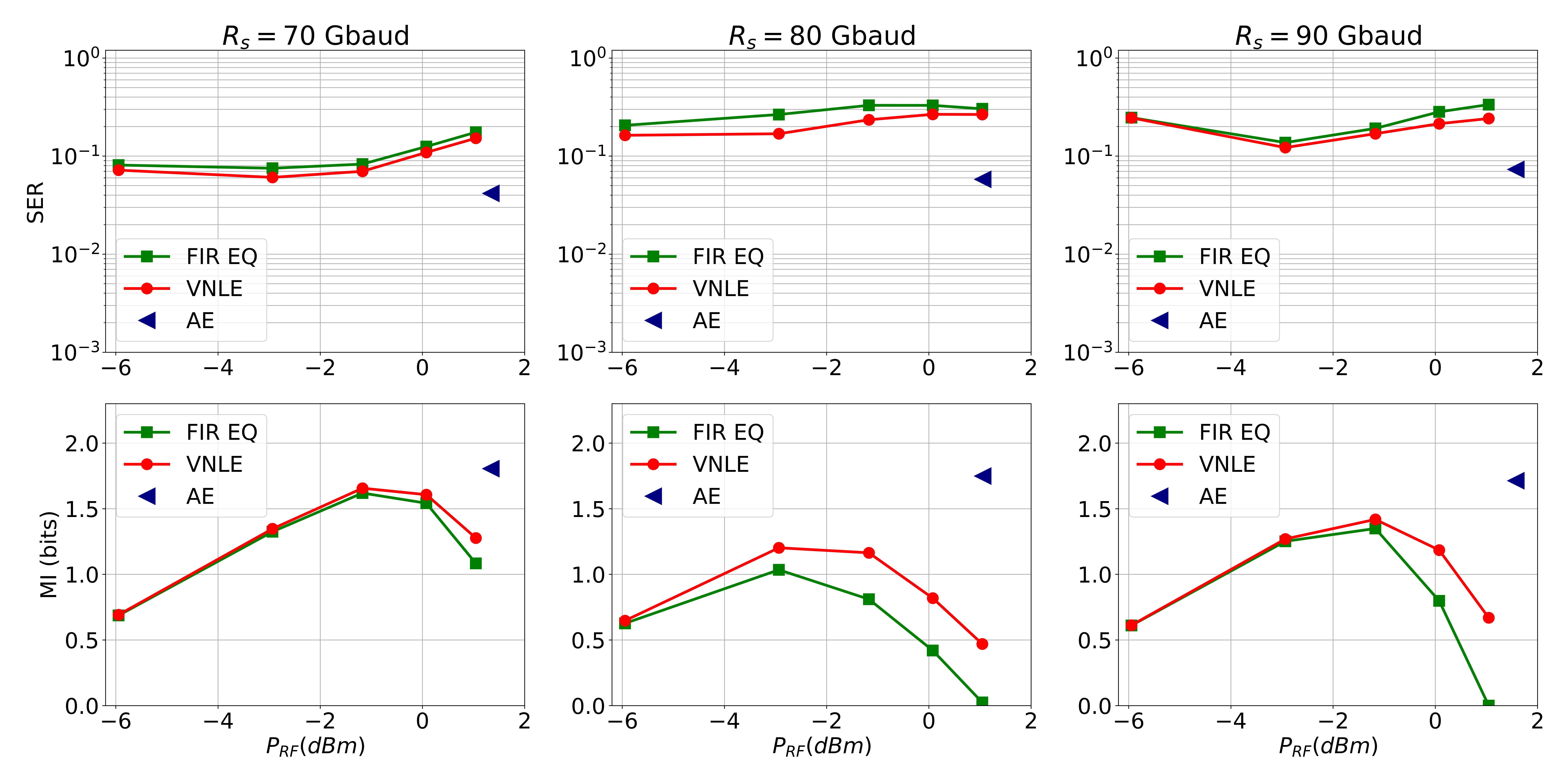}
    \caption{Link performance results in terms of symbol error rate (top) and mutual information (bottom) at symbol rates \{70,80,90\} Gbaud \{left, middle, right\}}
    \label{fig:results}
\end{figure*}

\section{Results}
\vspace{-2 mm}
The surrogate prediction results for each studied $R_s$ are depicted in \cref{fig:surr}. The surrogate is able to approximate $f_\mathrm{DML}$ with a relative test NRMSE of less than 1\% across the studied symbol rates. The AE results are contextualized with 2 benchmark RX-side equalization schemes, one based on a linear FIR EQ and another based on a second-order VNLE. All 3 models have a total of 20 memory taps (15 at DPD + 5 at EQ in the case of the AE, 20 linear for the FIR EQ and 20 linear and 20 nonlinear for the VNLE) in order to establish a balanced comparison. The benchmark schemes use equispaced 4PAM constellation with rectangular pulse shaping. Given that $I_\mathrm{bias}$ is not optimized in the benchmark setups, it is set to 30~mA (maximum bias level allowed to the AE) to minimize nonlinear waveform distortion. $I_\mathrm{pp}$ is swept between 4 and 20 mA in steps of 4 mA, yielding 5 different $P_{RF}$ levels. The benchmark schemes are trained using mean squared error between TX and RX 2-SpS waveforms, but SER and MI are calculated using maximum likelihood detection. The SER and MI metrics of the AE are obtained by evaluating its performance directly on the rate equations, in order to avoid any unfair advantage due to surrogate model inaccuracies. The results in \cref{fig:results} show the superior performance of the AE in the presence of strong signal impairments with respect to the benchmark schemes. Its advantage seems to be especially noticeable for higher $R_s$, where the nonlinear DML response and the CD-chirp interaction is most prominent. The high error rates in all three setups can be attributed to the small number of taps utilized given the intense distortion of the channel. A decision feedback equalizer could significantly improve error performance, but the non-differentiability of the hard decision operation does not allow joint optimization in the proposed setup. The FIR EQ and VNLE curves show a performance decay for $P_{RF}$ beyond -2 dBm. This is due to the enhanced nonlinear dynamics at high optical powers, that creates waveform distortion. At higher $R_s$, the bandwidth limitation and the chirp-CD interplay becomes more prominent, worsening the MI and SER performance even further compared to the 70 Gbaud case. The performance of the AE is also impacted at higher $R_s$, although mildly. The resilience of the AE setup to high symbol rates may suggest that E2E impairment compensation is specially powerful at compensating high ISI effects in direct detection systems. Another interesting result is the $I_\mathrm{bias}$ level delivered by the AE optimization, that varies between 15.4 and 15.9~mA for the 3 tested $R_s$. Given that the maximum allowed $I_\mathrm{bias}$ to the AE was of 30 mA, this may suggest that the AE is able to exploit the nonlinear behaviour to improve error performance, instead of suppressing it by maximizing $I_\mathrm{bias}$. Lower bias levels could also allow lower average transmitted power and energy per bit, a crucial advantage in the context of IM/DD systems.
\vspace{-5 mm}
\section{Conclusions}
We have proposed the use of end-to-end impairment compensation for directly modulated lasers using a novel chirp-aware surrogate model. We compare our approach to linear and nonlinear receiver-side equalization using the same number of memory taps in all cases. The proposed autoencoder setup is able to outperform the receiver-side equalization approaches throughout all the tested symbol rates, showing great potential in the impairment compensation of dispersion-limited directly modulated laser links. 

\clearpage
\section{Acknowledgements}
The Villum Fonden (VI-POPCOM VIL54486 and OPTIC-AI VIL29334) is acknowledged.

\defbibnote{myprenote}{%
Citations must be easy and quick to find. More precisely:
\begin{itemize}
    \item Please list all the authors. 
    \item The title must be given in full length. 
    \item Journal and conference names should not be abbreviated but rather given in full length.
    \item The DOI number should be added incl. a link.
\end{itemize}
}
\printbibliography

@ARTICLE{9874980,
  author={Guan, Shijian and Zhang, Yunshan and Zheng, Jilin and Su, Jingyou and Sun, Zhenxing and Lu, Linlin and Fang, Tao and Li, Lianyan and Xiao, Rulei and Shi, Yuechun and Chen, Xiangfei},
  journal={Journal of Lightwave Technology}, 
  title={{Modulation Bandwidth Enhancement and Frequency Chirp Suppression in Two-Section DFB Laser}}, 
  year={2022},
  volume={40},
  number={22},
  pages={7383-7389},
  doi={10.1109/JLT.2022.3203723}}

@ARTICLE{9720189,
  author={Diamantopoulos, Nikolaos-Panteleimon and Fujii, Takuro and Yamaoka, Suguru and Nishi, Hidetaka and Takeda, Koji and Tsuchizawa, Tai and Segawa, Toru and Kakitsuka, Takaaki and Matsuo, Shinji},
  journal={Journal of Lightwave Technology}, 
  title={{60 GHz Bandwidth Directly Modulated Membrane III-V Lasers on SiO2/Si}}, 
  year={2022},
  volume={40},
  number={10},
  pages={3299-3306},
  doi={10.1109/JLT.2022.3153648}}

@ARTICLE{10382548,
  author={Hernandez F., Sergio and Jovanovic, Ognjen and Peucheret, Christophe and Ros, Francesco Da and Zibar, Darko},
  journal={IEEE Photonics Technology Letters}, 
  title={{Differentiable Machine Learning-Based Modeling for Directly-Modulated Lasers}}, 
  year={2024},
  volume={36},
  number={4},
  pages={266-269},
  doi={10.1109/LPT.2024.3350993}}

@article{REZA2020102322,
title = {{Blind Nonlinearity Mitigation of 10G DMLs using Sparse Volterra Equalizer in IM/DD PAM-4 Transmission Systems}},
journal = {Optical Fiber Technology},
volume = {59},
pages = {102322},
year = {2020},
issn = {1068-5200},
doi = {10.1016/j.yofte.2020.102322},
author = {Ahmed Galib Reza and June-Koo Kevin Rhee},
}

@inproceedings{Srinivasan:22,
author = {Muralikrishnan Srinivasan and Jinxiang Song and Christian H\"{a}ger and Krzysztof Szczerba and Henk Wymeersch and Jochen Schr\"{o}der},
booktitle = {European Conference on Optical Communication (ECOC) 2022},
journal = {European Conference on Optical Communication (ECOC) 2022},
pages = {We2C.3},
title = {{Learning Optimal PAM Levels for VCSEL-based Optical Interconnects}},
year = {2022},
}

@inproceedings{Kottke:17,
author = {Christoph Kottke and Christoph Caspar and Volker Jungnickel and Ronald Freund and Mikel Agustin and Nikolay N. Ledentsov},
booktitle = {Optical Fiber Communication Conference},
journal = {Optical Fiber Communication Conference},
pages = {W4I.7},
title = {{High Speed 160 Gb/s DMT VCSEL Transmission Using Pre-equalization}},
year = {2017},
doi = {10.1364/OFC.2017.W4I.7},
}

@ARTICLE{8207825,
  author={},
  journal={IEEE Std 802.3bs-2017}, 
  title={{IEEE Standard for Ethernet - Amendment 10: Media Access Control Parameters, Physical Layers, and Management Parameters for 200 Gb/s and 400 Gb/s Operation}}, 
  year={2017},
  volume={},
  number={},
  pages={270},
  doi={10.1109/IEEESTD.2017.8207825}}

@INPROCEEDINGS{8553260,
  author={Tan, U. and Rabastc, O. and Adnet, C. and Ovarlez, J.-P.},
  booktitle={26th European Signal Processing Conference (EUSIPCO)}, 
  title={{A Sequence-Filter Joint Optimization}}, 
  year={2018},
  volume={},
  number={},
  pages={2335-2339},
  doi={10.23919/EUSIPCO.2018.8553260}}

\vspace{-4mm}

\end{document}